\documentstyle[12pt,cite,axodraw,epsfig]{article}
\newcommand{\be}{\begin{equation}}
\newcommand{\ee}{\end{equation}}
\newcommand{\etal}{{\em et al.}}
\newcommand{\eg}{{\em e.g.}}

\newcommand{\eqref}[1]{Eq.~(\ref{#1})}

\begin{document}

\def\be{\begin{equation}}
\def\bea{\begin{eqnarray}}
\def\ee{\end{equation}}
\def\eea{\end{eqnarray}}
                              \def\barr{\begin{array}}
                              \def\earr{\end{array}}
\def\dis{\displaystyle}
\def\eg{{\em e.g.}}
\def\etc{{\em etc.}}
\def\etal{{\em et al.}}

\def\ie{{\em i.e.}}
\def\viz{{\em viz.}}
\def\lsim{\:\raisebox{-0.5ex}{$\stackrel{\textstyle<}{\sim}$}\:}
\def\gsim{\:\raisebox{-0.5ex}{$\stackrel{\textstyle>}{\sim}$}\:}
                              \def\mev{\: \rm MeV} 
                              \def\gev{\: \rm GeV} 
                              \def\tev{\: \rm TeV} 
                              \def\pb {\: \rm pb}
                              \def\fb {\: \rm fb}
\def\gappeq{\mathrel{\rlap {\raise.5ex\hbox{$>$}}
            {\lower.5ex\hbox{$\sim$}}}}
\def\lappeq{\mathrel{\rlap{\raise.5ex\hbox{$<$}}
            {\lower.5ex\hbox{$\sim$}}}}
\def\ra{\rightarrow}
\def\mand{\qquad {\rm and} \qquad}
                              \def\ptsl{p_T \hspace{-1.1em}/\;}
                              \def\pslash{p \hspace{-0.6em}/\;}
                              \def\msl{m \hspace{-0.8em}/\;}
\def\rp{R\!\!\!/ _p}
\def\nut{\chi^0}
\def\cha{\chi^\pm}
\def\squ{\tilde q}
\def\slep{\tilde l}
\def\slashiii#1{\setbox0=\hbox{$#1$}#1\hskip-\wd0\hbox to\wd0{\hss\sl/\/\hss}}
\def\slashiv#1{#1\llap{\sl/}}
\newcommand{\rpv}{$\slashiii{R}$}
\begin{flushright}
{\large \tt hep-ph/0102180}
\end{flushright}

\vspace*{2ex}

\begin{center}
{\Large\bf A nonsupersymmetric resolution of the anomalous muon magnetic 
	   moment}

\vskip 15pt

{\sf Debrupa Chakraverty$^{1}$, Debajyoti Choudhury${^2}$ and Anindya Datta$^{3}$}

\vskip 5pt

{\footnotesize Harish Chandra Research Institute, 
Chhatnag Road, Jhusi, Allahabad 211 019, India. \\
E-mail: $^1$rupa@mri.ernet.in, $^2$debchou@mri.ernet.in, 
        $^3$anindya@mri.ernet.in}\\

\vskip 15pt
{\Large\bf Abstract}

\end{center}

\begin{quote}
The recent result from the E821 experiment at BNL on 
the anomalous magnetic moment of the muon 
shows a distinct discrepancy with the Standard
Model predictions. We calculate the additional 
correction that the anomalous magnetic moment receives 
in a model with scalar leptoquarks. We find that such models can 
account for the deviation from the SM value even for small 
leptoquark couplings. 
\end{quote}
\vskip 20pt


\vskip 20pt
\hspace*{0.65cm}

\def\baselinestretch{1.0}



\renewcommand{\thefootnote}{\arabic{footnote}}
\setcounter{footnote}{0}

The recent measurement~\cite{BNL_new} of the magnetic dipole moment of
the muon has set off a flurry of excitement amongst
theorists~\cite{marciano,lane,kane,feng,baltz}.  The unprecedented
precision achieved  seems to imply that the experimentally 
observed value
disagrees with the Standard Model (SM) expectations at more than $2.6
\sigma$ level. If this discrepancy is to be accepted at its face
value, it seems to indicate the presence of new physics just round the
corner. The exact nature of this new physics is a matter of intense
speculation though. 
Since it is difficult to accommodate this deviation within a 
large class of models such as those with 
left-right symmetry or anomalous gauge boson 
couplings~\cite{marciano} or a world with large extra 
dimensions~\cite{extrad}, it is natural that most 
practitioners favour supersymmetry as a 
solution~\cite{kane,feng,baltz,old_susy}.  Even within the family
of supersymmetric models, certain classes are less favoured than
others, an example of the former 
being afforded by scenarios with anomaly mediated
supersymmetry breaking~\cite{feng,utpal}.  
The favoured models, on the other
hand, require that the superpartners be relatively light and within
reach of the next run at the Fermilab Tevatron. It, thus, is of great
importance to examine other possible extensions of the SM that can
accommodate the measured value of the muon magnetic moment. In this
article, we argue that scalar leptoquarks offer a perfectly viable
solution to the problem.

Leptoquarks, as the name suggests, are particles that couple to a
current comprising of a lepton and a quark. Arising naturally in many
models with extended gauge symmetry (including but not limited to
grand unification), they have been studied extensively in the
literature~\cite{leptoquarks,sacha}. 
In this article, we shall confine ourselves to a discussion of 
scalar leptoquarks 
\footnote{While vector leptoquarks corresponding to a gauge 
	symmetry would tend to be superheavy, non-gauged
	vector particles generically imply a lack of renormalizability 
	in the theory.}, 
a class that includes the squarks in a 
$R$-parity violating supersymmetric model. 
Relevance to the current context demands that these states
be relatively light (certainly $\lsim {\cal O}(1 \tev)$ or so).
While within supersymmetric models, their mass can be 
protected by nonrenormalizability theorems, in a generic model, 
additional discrete symmetries may ensure this. 
On account of the leptoquark coupling violating both baryon 
and lepton numbers, it might appear, at first sight, 
that such a light state might lead to rapid
proton decay.  However, it is easy to see that the proton would decay 
only if the said 
leptoquark either couples to a two-quark current or mixes
with another scalar which does so. Phenomenological consistency thus
demands that any such two-quark coupling (or mixing) 
be severely suppressed, 
a requirement that can be easily satisfied in most models.

\begin{table}[h]
\begin{center}
\begin{tabular}{|c|c|c|}
\hline
& &  \\[-2ex]
 Leptoquark Type  &  Coupling & $SU(3)_c\times SU(2)_L\times U(1)_Y$\\
& &  \\[-2ex]
 \hline
 \hline
& &  \\[-2ex]
 $\Phi_1$ & $\left[\lambda_{ij}^{(1)} {\bar Q}_{L j} e_{R i} 
 + {\tilde \lambda}_{ij}^{(1)} {\bar u}_{R j} L_{Li}\right]\Phi_1
 $ & 
	($3$, $2$, ${7\over 3}$)\\ 
& &  \\[-2ex]
 \hline
& &  \\[-2ex]
 $\Phi_2$ & $\lambda_{ij}^{(2)} {\bar Q}_{L j}^c L_{L i} \Phi_2$ & 
	(${\bar 3}$, $3$, ${2\over 3}$)\\ 
& &  \\[-2ex]
 \hline
& &  \\[-2ex]
& &  \\[-2ex]
 $\Phi_3$ & $\left[\lambda_{ij}^{(3)} {\bar Q}_{L j}^c L_{L i} 
 + {\tilde \lambda}_{ij}^{(3)} {\bar u}_{R j}^c e_{Ri}\right]\Phi_3
 $ & 
	(${\bar 3}$, $1$, ${2\over 3}$)\\ 
& &  \\[-2ex]
 \hline
& &  \\[-2ex]
 $\Phi_4$ & $\lambda_{ij}^{(4)} {\bar d}_{R j} L_{L i} \Phi_4$ & 
	($3$, $2$, ${1\over 3}$)\\ 
& &  \\[-2ex]
 \hline
& &  \\[-2ex]
 $\Phi_5$ & $\lambda_{ij}^{(5)} {\bar d}_{R j}^c e_{R i} \Phi_5$ & 
	(${\bar 3}$, $1$, ${8\over 3}$)\\ 
 \hline
\end{tabular}
\end{center}
    \caption{\em Gauge quantum numbers and Yukawa couplings 
	         of scalar leptoquarks $(Q_{em} = T_3 + \frac{Y}{2})$.
            }
	\label{table:qnos}
\end{table}

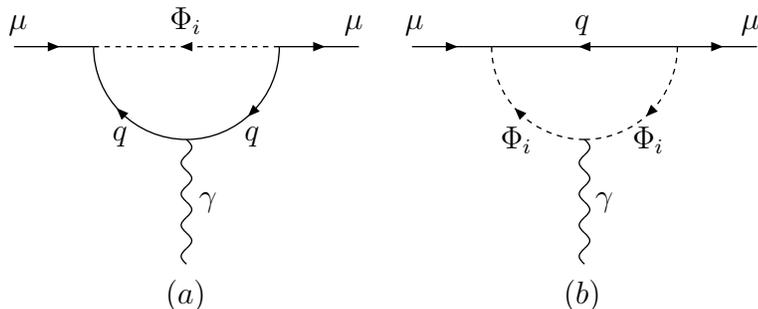
\begin{figure}[h]
\begin{center}
\begin{picture}(300,108)(0,-8)
\ArrowLine(0,100)(30,100)
\Text(5,107)[rb]{$\footnotesize{\mu}$}
\ArrowLine(100,100)(130,100)
\Photon(65,65)(65,18){2}{4}
\Text(70,40)[lb]{$\footnotesize{\gamma}$}
\Text(125,107)[lb]{$\footnotesize{\mu}$}
\DashArrowLine(100,100)(30,100){2}
\Text(65,107)[cb]{$\footnotesize{\Phi_i}$}
\ArrowArcn(65,100)(35,270,180)
\ArrowArcn(65,100)(35,0,270)
\Text(40,70)[ct]{$\footnotesize{q}$}
\Text(90,70)[ct]{$\footnotesize{q}$}
\Text(65,12)[ct]{$\footnotesize{(a)}$}
\ArrowLine(150,100)(180,100)
\Text(155,107)[rb]{$\footnotesize{\mu}$}
\ArrowLine(250,100)(180,100)
\ArrowLine(250,100)(280,100)
\Photon(215,65)(215,18){2}{4}
\Text(220,40)[lb]{$\footnotesize{\gamma}$}
\Text(275,107)[lb]{$\footnotesize{\mu}$}
\Text(215,107)[cb]{$\footnotesize{q}$}
\DashArrowArcn(215,100)(35,270,180){2}
\DashArrowArcn(215,100)(35,0,270){2}
\Text(240,70)[ct]{$\footnotesize{\Phi_i}$}
\Text(190,70)[ct]{$\footnotesize{\Phi_i}$}
\Text(215,12)[ct]{$\footnotesize{(b)}$}
\end{picture} 
\end{center}
\vspace*{-7ex}
      \caption{\em Feynman diagrams that determine the leptoquark contribution 
		    to $a_\mu$.}
	\label{fig:feyn_diag}
\end{figure}

Rather than confine ourselves to a particular scenario, let us start 
by considering a generic scalar leptoquark. In Table~\ref{table:qnos}, 
we list all the possible states that can couple to a SM lepton
and quark pair. Confining ourselves to terms involving the muon field, 
the relevant part of the Lagrangian can be parametrized
as 
\be
{\cal L}_{\rm Yukawa} = 
\bar q_i (\lambda_L^{(A)} P_L + \lambda_R^{(A)} P_R) \mu \: \phi_{A} + H.c.,
	\label{Lagrangian}
\ee
where $\phi_{A}$ is one of the leptoquarks in Table~\ref{table:qnos}. 
For a given $\phi_{A}$, the structure of the chiral couplings 
$\lambda_{L, R}$ is determined by
its quantum numbers and can be easily read off from Table~\ref{table:qnos}.
 It should be noted that, in
eq.(\ref{Lagrangian}), the field $q_i$ may refer either to one of the
usual SM quarks or its charge conjugate. The above Lagrangian leads to
diagrams as in Fig.~\ref{fig:feyn_diag} that may contribute to the
muon dipole magnetic moment $g_\mu$. As the corresponding effective
operator
\[
\frac{e g_\mu}{2 m_\mu}
 \bar \mu \sigma_{\alpha \beta} \mu \: F^{\alpha \beta}
\]
is a chirality changing one, the corrections due to the diagrams of
Fig.~\ref{fig:feyn_diag} would be proportional to some fermion mass.  For
a generic diagram, this chirality flip can occur either in the
external legs or along the internal lines, the relative sizes of the
contributions being determined by the chirality structure of the
leptoquark coupling as well as the mass of the internal quark. It is
easy to see, that, for a term proportional to $m_q$ to be present, one
needs to have {\em both} of $\lambda_{L, R}$ in eq.(\ref{Lagrangian})
to be nonzero, a condition that 
can be satisfied only for $\Phi_{1,3}$. As we shall see later, the size of
the discrepancy implies that only such leptoquarks are relevant in
this context.


Defined as $a_\mu \equiv (g_\mu - 2)/ 2$, the anomalous magnetic 
dipole moment for the positively charged muon has been measured by the 
E821 Collaboration~\cite{BNL_new}, to be 
\be
	a_\mu^{\rm exp} = (11\ 659\ 202\, \pm 14 \pm 6) \times 10^{-10}
\ee
where the first uncertainty is statistical and the second systematic. 
This accuracy (1.3 ppm) has been achieved solely on the basis 
of the 1999 data.  Analysis of last year's data, currently 
underway, should reduce the error to $\sim 7 \times 10^{-10}$
(0.6 ppm), with the final targeted accuracy being $4 \times 10^{-10}$
(0.35 ppm)~\cite{future}. When compared to the SM 
value~\cite{smvalue}
\be
    a_\mu^{\rm SM} = (11\ 659\ 159.7 \pm 6.7) \times 10^{-10} \ ,
\ee
the new world average leads to a $2.6 \sigma$ discrepancy, {\em viz.}
\be
   \delta a_\mu \equiv a_\mu^{\rm exp} - a_{\mu}^{\rm {SM}} = (42.6 \pm 16.5)
			 \times	10^{-10} \ .
\ee
It might seem that the absolute magnitude of the deviation is small and 
that it should be easy to accommodate it by extending the SM. However, 
it should be noted that the bulk of the corrections are accounted 
for by QED loops~\cite{smvalue}
with hadronic vacuum polarization coming in a distant 
second~\cite{qcd_corr}. Standard weak interactions account 
for~\cite{weak_corr} only 
$(15.2 \pm 0.4) \times 10^{-10}$, a contribution significantly smaller 
than the size of the discrepancy. 

Reverting back to the diagrams of Fig.~\ref{fig:feyn_diag}, we see that, 
to ${\cal O}(m_\mu / m_q, m_\mu / m_\phi)$,  
the leptoquark contribution to $a_\mu$ is given by
\be
\barr{rcl}
a_\mu^{(\phi)} & = & \dis {- N_c m_\mu \over {8 \pi^2 m_{\phi}^2}}
	\Bigg[ m_q \lambda_L \lambda_R 
		\left\{ Q_{\phi} f_1(x) + Q_{q} f_2(x) \right\}
	\\[1.5ex]
	& & \dis \hspace*{3em}
	      + m_\mu ( \lambda_L^2 + \lambda_R^2) 
		\left\{ Q_{\phi} f_3(x) + Q_{q} f_4(x) \right\}
	\Bigg]
\earr
	\label{amu_expr}
\ee
where $x \equiv m_q^2 / m_\phi^2$ and 
\be
\barr{rcl}
f_1(x) & = & \dis {1\over {2 (1-x)^3}} [1 -x^2 + 2 x \ln x]  \\[1.5ex]
f_2(x) & = & \dis {1\over {2 (1-x)^3}} [3 -4 x +x^2 + 2  \ln x] \\[1.5ex]
f_3(x) & = & \dis  {1\over {12 (1-x)^4}} 
	[6 x^2 \ln x - 2 x^3 - 3 x^2 + 6 x -1]\\[1.5ex]
f_4(x) & = & \dis  {1 \over {12 (1-x)^4}} [6 x \ln x  +2 + 3 x - 6 x^2 + x^3] 
  \ .
\earr
\ee
In eq.(\ref{amu_expr}), $N_c = 3$ is the color factor and we have 
suppressed both the generation indices and the superscript denoting 
the leptoquark type.

\begin{figure}[ht]
\centerline{
\epsfxsize=6.5cm\epsfysize=6.0cm
                     \epsfbox{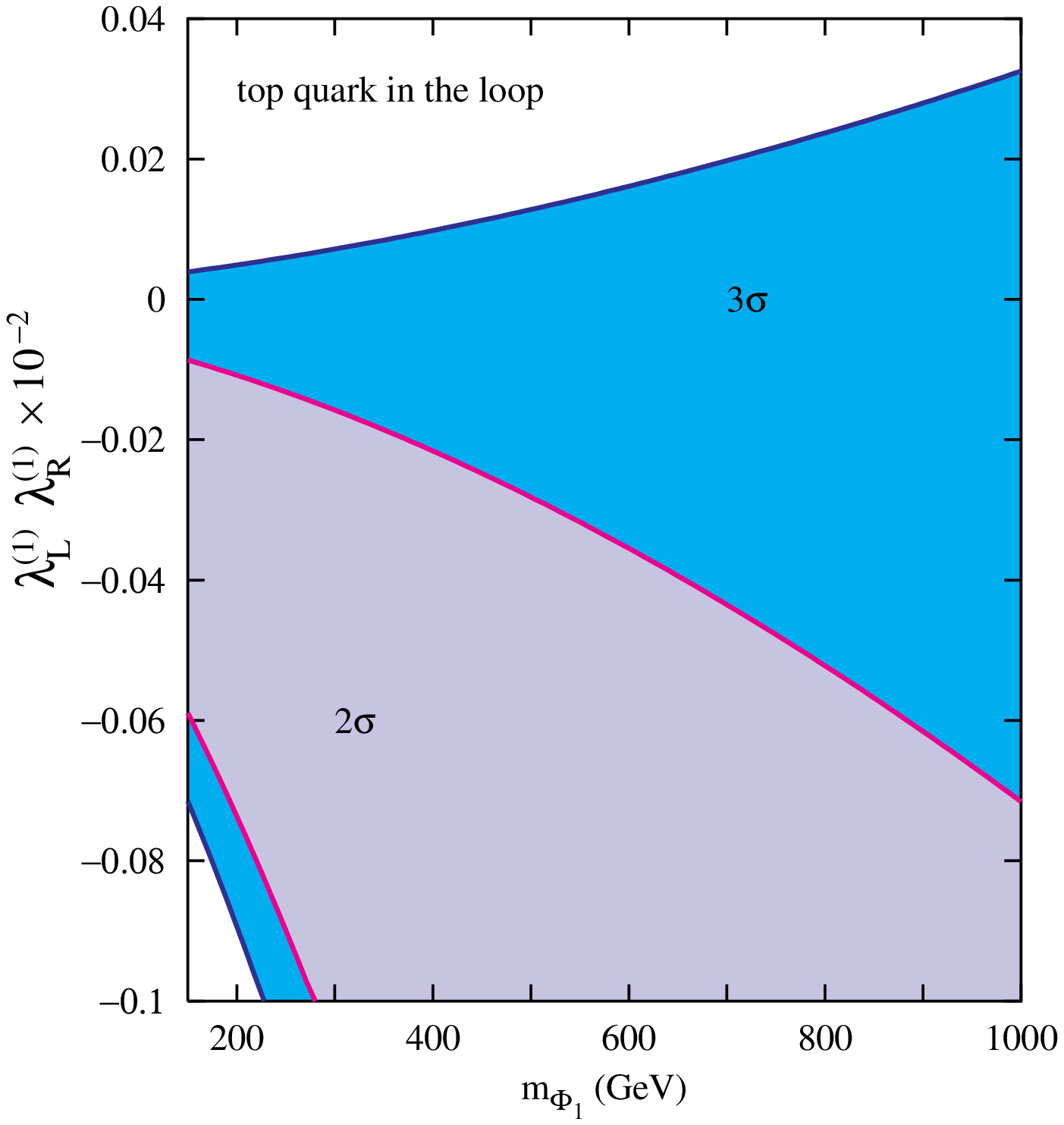}
        \hspace*{-1ex}
\epsfxsize=6.5cm\epsfysize=6.0cm
                     \epsfbox{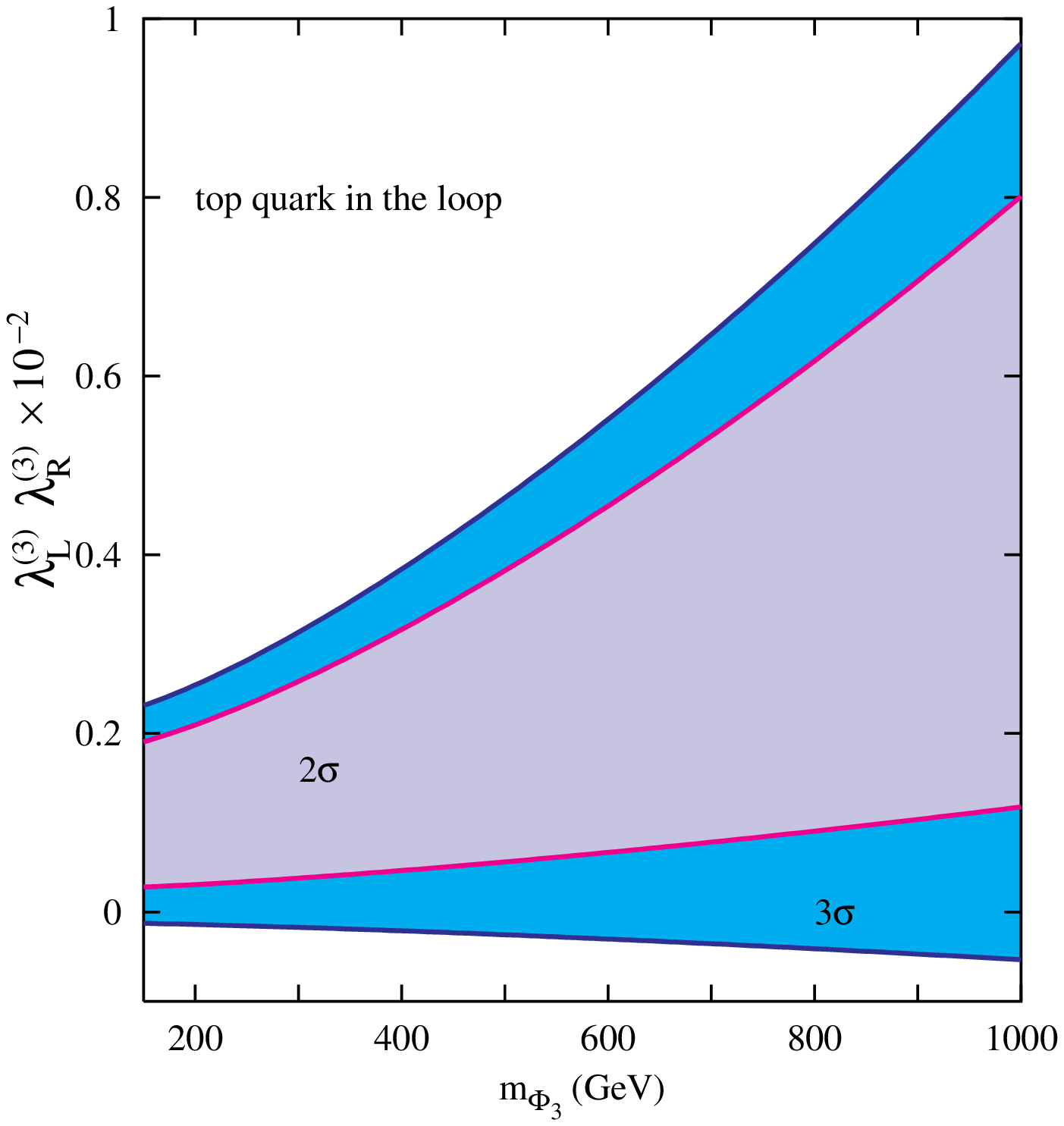}
}

\caption{\em The region of the parameter space consistent with the 
	$a_\mu$ measurement for {\em (a)} $\Phi_1$; and 
	{\em (b)} $\Phi_3$. It has been assumed that the only 
	two non-zero couplings are those involving the top quark. 
	 The lightly shaded 
	area agrees with the data at $2 \sigma$ level, whereas 
	the encompassing darker region agrees at $3 \sigma$.
	}
\label{fig:contour}
\end{figure}

As we have already mentioned, an ``explanation'' of $\delta a_\mu$
needs the term proportional to $m_q$ to be nonzero. In other words, we
need the leptoquark field to be of the types $\Phi_{1,3}$ and the
quark to be of the second or third generation. For the sake of
concreteness, let us, for the time being,  concentrate on the coupling
to the top quark. In Fig.~\ref{fig:contour}, we describe the region of
the parameter space that is consistent with the data at different
levels of confidence. Let us concentrate first on the case of $\Phi_1$. 
Since we can safely neglect the term proportional to $m_\mu$, 
the leptoquark contribution to $\delta a_\mu$ is essentially proportional
to the product $\lambda_R \lambda_L$. While the 
function $f_1(x)$ is essentially positive (except for very small $x$), 
$f_2(x)$ is always negative and larger in magnitude compared to 
$f_1(x)$. Given the quantum numbers of 
$\Phi_1$, this results in $a_\mu^{(\Phi_1)}$ having a sign opposite to that 
of the product $\lambda_R \lambda_L$. Since $\delta a_\mu \gsim 0$, this 
implies that a negative value of this product is preferred. 
For $\Phi_3$, the situation is the opposite. Here, $a_\mu^{(\Phi_3)}$
has the same sign as the product, and consequently, the curves look 
upside down compared to those for $\Phi_1$. Note, however, that the 
scale on the product axis is quite different. This difference owes itself to 
the amount of cancellation between the contributions of the two 
Feynman diagrams in Fig.~\ref{fig:feyn_diag}. 
Since the cancellation is more pronounced for $\Phi_3$ than for $\Phi_1$,
typically smaller values of $a_\mu^{(\phi)}$ result. Consequently, an 
agreement with the data requires larger values for the couplings.

It appears, then, that the presence of either of $\Phi_{1,3}$ 
can serve to explain $\delta a_\mu$. However, before we make
such a claim, it is contingent upon us to examine the existing
constraints on such a scenario. We proceed to do this next.

At the Tevatron, leptoquark production is dominated by strong
interactions and proceeds primarily through $q \bar q$
fusion. Subsequent decays into a quark (jet)-lepton pair has been
extensively looked for by both the CDF and the
D0~\cite{Tev_lepto} collaborations. Nonobservation of such signals
imply that a leptoquark decaying 
entirely\footnote{For branching fractions less than unity, the 
	bounds are understandably weaker}
into a light quark and a $e/\mu$ must be heavier than approximately
$230 \gev$.  For a leptoquark decaying primarily into the top, the
bounds would be weakened 
somewhat\footnote{To the best of our knowledge, this analysis 
	has not yet been presented by either of the collaborations.}. 
But what about the $I_3 = -1/2$ partner, which would be 
produced as abundantly\footnote{Consideration of the $\rho$-parameter 
	demands that the mass splitting between states in a multiplet 
	bed tiny.} and 
decays into a ($b + \nu_\mu$)-pair leading to an 
additional signal. The experimental efficiency for
this channel is low though and the corresponding bound~\cite{Tev_lepto}
is only $m_\phi \gsim 150 \gev$. In fact, even combining the two individual
bounds is unlikely to result in a constraint stronger than that 
for the `first-' or `second'-generation leptoquark as 
described above. 

Apart from collider search experiments, low-energy data could,
conceivably, also be used to constrain the parameter spaces for
individual leptoquarks~\cite{sacha}. However, it is easy to see that
the couplings to the top-quark are unconstrained by any such data.
In fact, the strongest bound on such couplings come from LEP 
data~\cite{gg_ellis_sridhar} on $Z \ra \ell \bar \ell$. Consistency 
with data typically requires that $\lambda \lsim g_{\rm Weak}$, a 
constraint that is easily satisfied by the entire parameter space 
of Fig.~\ref{fig:contour}. 

\begin{figure}[ht]
\centerline{
\epsfxsize=6.5cm\epsfysize=6.0cm
                     \epsfbox{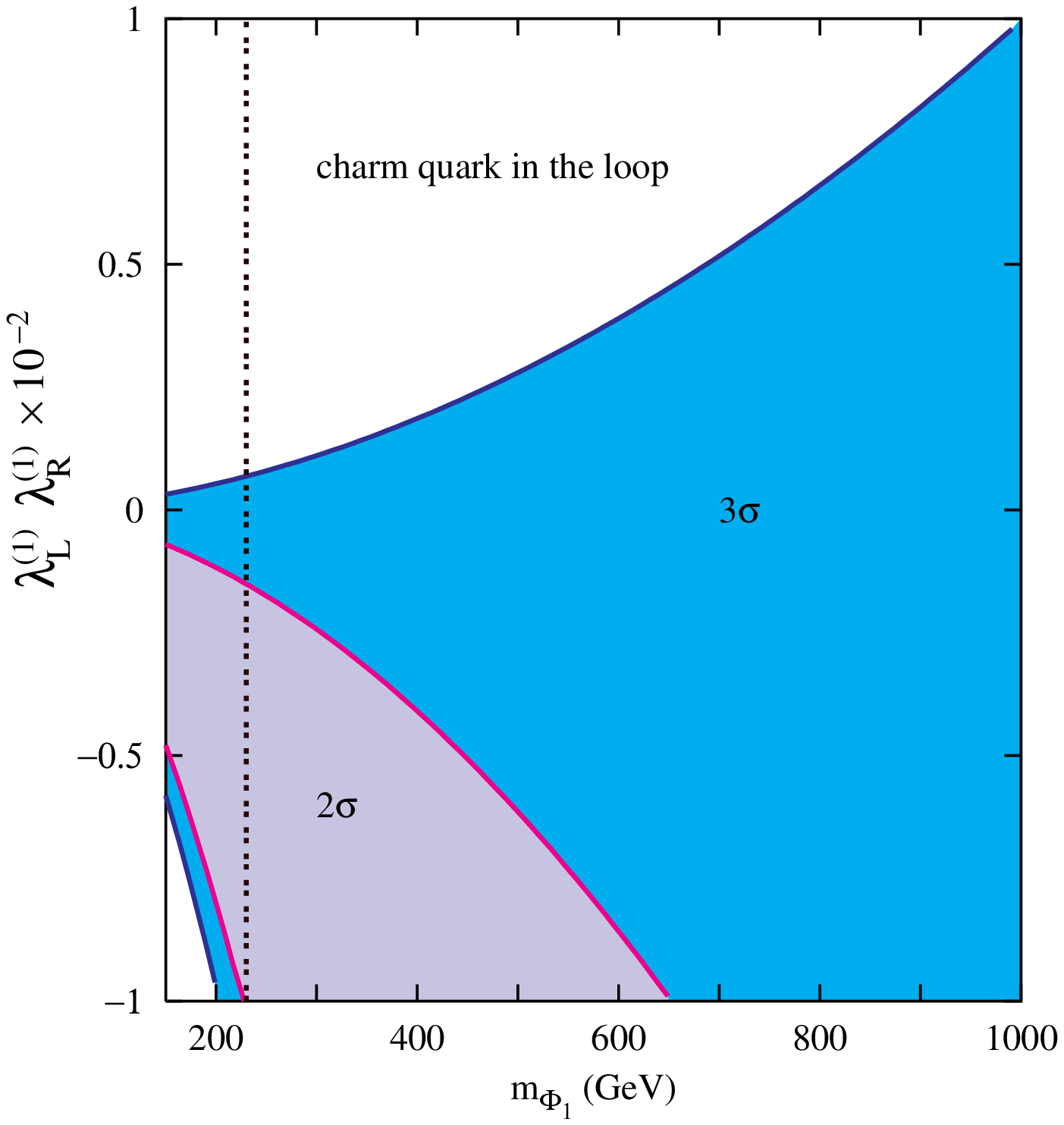}
        \hspace*{-1ex}
\epsfxsize=6.5cm\epsfysize=6.0cm
                     \epsfbox{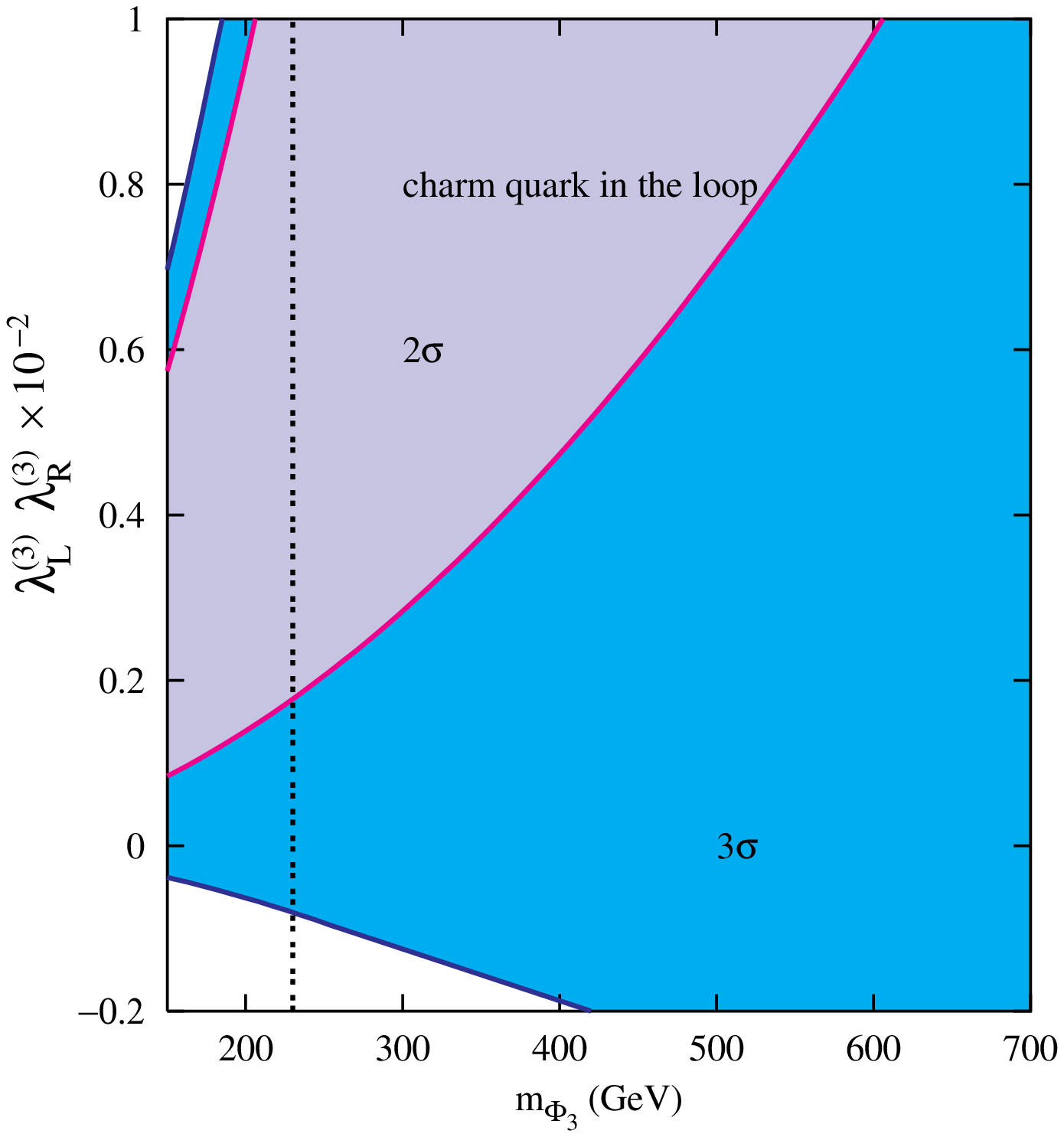}
}

\caption{\em As in Fig.~\protect\ref{fig:contour}, but for 
	couplings to the charm quark instead. 
	The region to the left of the vertical lines are 
	ruled out by direct search experiments at the 
	Tevatron~\protect\cite{Tev_lepto}.
	}
\label{fig:contour_c}
\end{figure}
Having established that a $\mu t \phi$ coupling 
can explain $\delta a_\mu$ while respecting all other known constraints, 
let us now turn to another possibility, namely $q = c$. 
The suppression 
$m_c / m_t$ immediately springs to mind. This, however, is ameliorated, 
to a significant extent, by the behaviour of $f_2 (x)$ as $x \ra 0$. 
The corresponding results are exhibited in Fig.~\ref{fig:contour_c}. 
Keeping in mind possible cancellation between terms, one might 
be tempted to question the neglect of the terms proportional to $m_\mu$. 
We have checked explicitly though that these continue to be numerically 
insignificant. Expectedly, somewhat larger values of the couplings 
are required although the suppression factor is not as large as $m_t / m_c$.
Still, are such values of the couplings allowed, especially 
by low energy phenomenology? A search through literature yields 
nothing, except for a very weak constraint~\cite{sacha} from 
old measurements of $a_\mu$ itself! Analysing all possible meson decays 
wherein the $\mu c \Phi_{1,3}$ couplings could play a role, we find 
that the most significant constraint emanates from the 
helicity-suppressed decay $D_s^+ \ra \mu^+ \nu_\mu$. If both 
$\lambda_{L, R}$ be nonzero, the leptoquark contribution to the 
decay amplitude is no longer mass suppressed, and the 
branching fraction reads 
\begin{eqnarray}
Br(D_s^+ \ra \mu^+ {\nu_\mu}) & = & {1 \over { 64 \pi}}  
f_{D_s}^2 m_{D_s}^3 \tau_{D_s} \left\vert {{4 G_f V_{cs}} 
\over \sqrt{2}} {m_\mu \over m_{D_s}} 
  - {{\lambda_L \lambda_R} \over {2 m_{\phi}^2}} \right\vert^2 \nonumber\\
& & \left( 1- {m_\mu^2 \over m_{D_s}^2}\right)^2
\end{eqnarray}
Comparing to the experimental number 
$Br(D_s^+ \ra \mu^+ {\nu_\mu}) = (4.6 \pm 1.9) \times 10^{-3}$~\cite{PDG},
leads, at the $2 \sigma$ level, to the constraint
\be \dis 
	-0.009 < 
	\lambda_L \lambda_R \: \left( \frac{m_\phi}{100 \gev} \right)^2
	< 0.078 \ .
\ee
In Fig.~\ref{fig:contour_c}, the lower limit would translate to a 
small parabolic curve at the extreme bottom left corner, a region 
already ruled out by direct searches~\cite{Tev_lepto}. The upper 
limit lies beyond the scale of the plot. It can thus be argued that,
the direct search limit is the only relevant constraint 
for the part of the parameter space consistent with the $a_\mu$ 
measurement. 

Having seen that even couplings with the charm-quark can be
instrumental in explaining $\delta a_\mu$, it is tempting to ask if a
similar result obtains for the up-quark as well. In this case though,
the growth in $f_2(x)$ cannot compensate enough for the smallmess of
$m_u$, and, for moderate values of the couplings, the leptoquark
contribution is too small to be of relevance. 

In summary, we have investigated the corrections  that a relatively light 
scalar leptoquark could wrought in coupling of the muon to the photon. 
We find that such a particle can indeed serve to reconcile the 
newly measured value of the magnetic dipole moment 
with theory. However, the leptoquark needs to have both left-handed 
and right-handed couplings to the muon field. This limits us to two 
species of leptoquarks out of a possible five. Experimental data 
prefers that these couple to the second or third generation quarks. 
The required magnitude of the couplings is on the smaller side and 
in consonance with all known bounds. The expected reduction in the 
error would serve to further constrain the region in parameter space
allowed by the current data.

\newcommand{\plb}[3]{{Phys. Lett.} {\bf B#1} (#3) #2}                  %
\newcommand{\prl}[3]{Phys. Rev. Lett. {\bf #1} (#3) #2}        %
\newcommand{\rmp}[3]{Rev. Mod.  Phys. {\bf #1} (#3) #2}             %
\newcommand{\prep}[3]{Phys. Rep. {\bf #1} (#3) #2}                     %
\newcommand{\rpp}[3]{Rep. Prog. Phys. {\bf #1} (#3) #2}             %
\newcommand{\prd}[3]{{Phys. Rev.}{\bf D#1} (#3) #2}                    %
\newcommand{\np}[3]{Nucl. Phys. {\bf B#1} (#3) #2}                     %
\newcommand{\npbps}[3]{Nucl. Phys. B (Proc. Suppl.) 
           {\bf #1} (#3) #2}                                           %
\newcommand{\sci}[3]{Science {\bf #1} (#3) #2}                 %
\newcommand{\zp}[3]{Z.~Phys. C{\bf#1} (#3) #2}                 %
\newcommand{\mpla}[3]{Mod. Phys. Lett. {\bf A#1} (#3) #2}             %
\newcommand{\astropp}[3]{Astropart. Phys. {\bf #1} (#3) #2}            %
\newcommand{\ib}[3]{{\em ibid.\/} {\bf #1} (#3) #2}                    %
\newcommand{\nat}[3]{Nature (London) {\bf #1} (#3) #2}         %
\newcommand{\nuovocim}[3]{Nuovo Cim. {\bf #1} (#3) #2}         %
\newcommand{\yadfiz}[4]{Yad. Fiz. {\bf #1} (#3) #2 [English            %
        transl.: Sov. J. Nucl.  Phys. {\bf #1} #3 (#4)]}               %
\newcommand{\philt}[3]{Phil. Trans. Roy. Soc. London A {\bf #1} #2  
        (#3)}                                                          %
\newcommand{\hepph}[1]{(electronic archive:     hep--ph/#1)}           %
\newcommand{\hepex}[1]{(electronic archive:     hep--ex/#1)}           %
\newcommand{\astro}[1]{(electronic archive:     astro--ph/#1)}         %

\end{document}